\documentclass[iop]{emulateapj}
\usepackage{xcolor}
\usepackage{graphicx}
\usepackage{adjustbox}
\usepackage{rotating}

\newcommand{\kms}{$\mathrm{\,km\,s}^{-1}$ }

\shorttitle{Kinematics of Halo Cepheids}
\shortauthors{Wallerstein \& Farrell}

\begin{document}

\title{Kinematics of Type II Cepheids of the Galactic Halo}

\author{George Wallerstein}
\author{Elizabeth~M. Farrell}
\affil{Deptartment of Astronomy, University of Washington, Seattle, WA 98195}

\begin{abstract}

In a step toward understanding the origin of the Galactic Halo, we have reexamined Type II Cepheids (T2C) in the field with new input from the second data release (DR2) of {\it Gaia}. For 45 T2C with periods from 1 to 20 days, parallaxes, proper motions, and [Fe/H] values are available for 25 stars. Only 5 show [Fe/H] $\textless$ --1.5, while the remaining stars show thick disk kinematics and [Fe/H] $\textgreater$ --0.90. We have compared the T2C stars of the field with their cousins in the globular clusters of the Halo and found that the globular clusters with T2C stars show metallicities and kinematics of a pure Halo population. The globulars may have formed during the overall collapse of the Galaxy while the individual thick disk T2C stars may have been captured from small systems that self-enriched prior to capture. The relationship of these two populations to the micro-galaxies currently recognized as surrounding the Galaxy is unclear.

\end{abstract}


\section{Introduction}
\label{sec:intro}

The concept of stellar populations was introduced by \citet{Baade1944}, who called attention to the differences between stars in the spiral arms of galaxies as compared to galactic halos and bulges. At the Vatican Conference (\citealt{OConnell1958}), the concepts of halo, thick disk, thin disk, and spiral arms were specified. The Halo was defined by the globular clusters and stars whose motions carried them to large distances above the Galactic Plane. Metallicity studies by \citet{Chamberlain1951}, \citet{Schwarzschild1957}, and \citet{Helfer1959} showed that both individual stars and globular clusters of the halo are deficient in metals by factors as large as 100. Subsequent surveys of individual stars and integrated light of globulars showed a substantial range in metallicity (\citealt{Deutsch1955}, \citealt{Kinman1959}). The assignment of a star, or a group of stars, to a specific population involves both the location and kinematics of the star or group. It is preferable to employ metallicity as a property of a population rather than as a defining parameter.  

Once the difference between Classical Cepheids and Cepheids in globular clusters had been recognized by \citet{Baade1956}, the T2C stars seemed to be tracers of the Halo. At that time, the T2C stars were mostly too faint for high dispersion analysis of their metallicity, but their radial velocities could be measured. Using their velocities and positions, \citet{Woolley1966} concluded that most, but not all, of the T2C stars at substantial galactic latitudes belonged to the thick disk population rather than to the Halo population and its globular cluster T2C stars. T2C stars are also found in the Large and Small Magellanic Clouds (\citealt{Soszynski2008, Soszynski2010}), but their velocities and metallicities have not been evaluated because of their faintness. In this paper, we present kinematic and metallicty data for individual T2C stars in the field, and compare their properties with T2C stars in globular clusters.

\section{Observations}
\label{sec:obs}

The second data release of {\it Gaia} provides a substantial upgrade in the avaiable data for T2C stars in the Galactic Halo. Over the 50 years since Woolley's paper, many Cepheids have been found at substantial distances above the Galactic Plane. In the General Catalogue of Variable Stars (\citealt{Samus2001}), they are listed as CW stars. The Catalogue of H.~C. \citet{Harris1985}, and new identifications by \citet{Schmidt2003, Schmidt2003b, Schmidt2004}, have provided a substantial database for T2C stars in the field. Metallicities have been provided by \citet{Maas2007}, \citet{Schmidt2003}, \citet{Harris1981}, \citet{Kovtyukh2018b}, and \citet{Lemasle2015}. So as to avoid the Galactic Bulge, we have excluded from our list stars with a galactic longitude between 350 and 10 degrees, and with a galactic latitude less than $\pm$ 10 degrees. Only the stars whose parallax exceeds the probable error of their parallax by a factor of at least 5 have been included.\footnote{There is no simple way to balance the need to include as much data as 
possible without including data with an undesireable level of uncertainty. For a discussion of parallax errors see \citet{Luri2018}, \citet{Bailer2018}, as well as \citet{Lutz1973}.} 

\begin{figure*}
\includegraphics[width=0.75\textwidth]{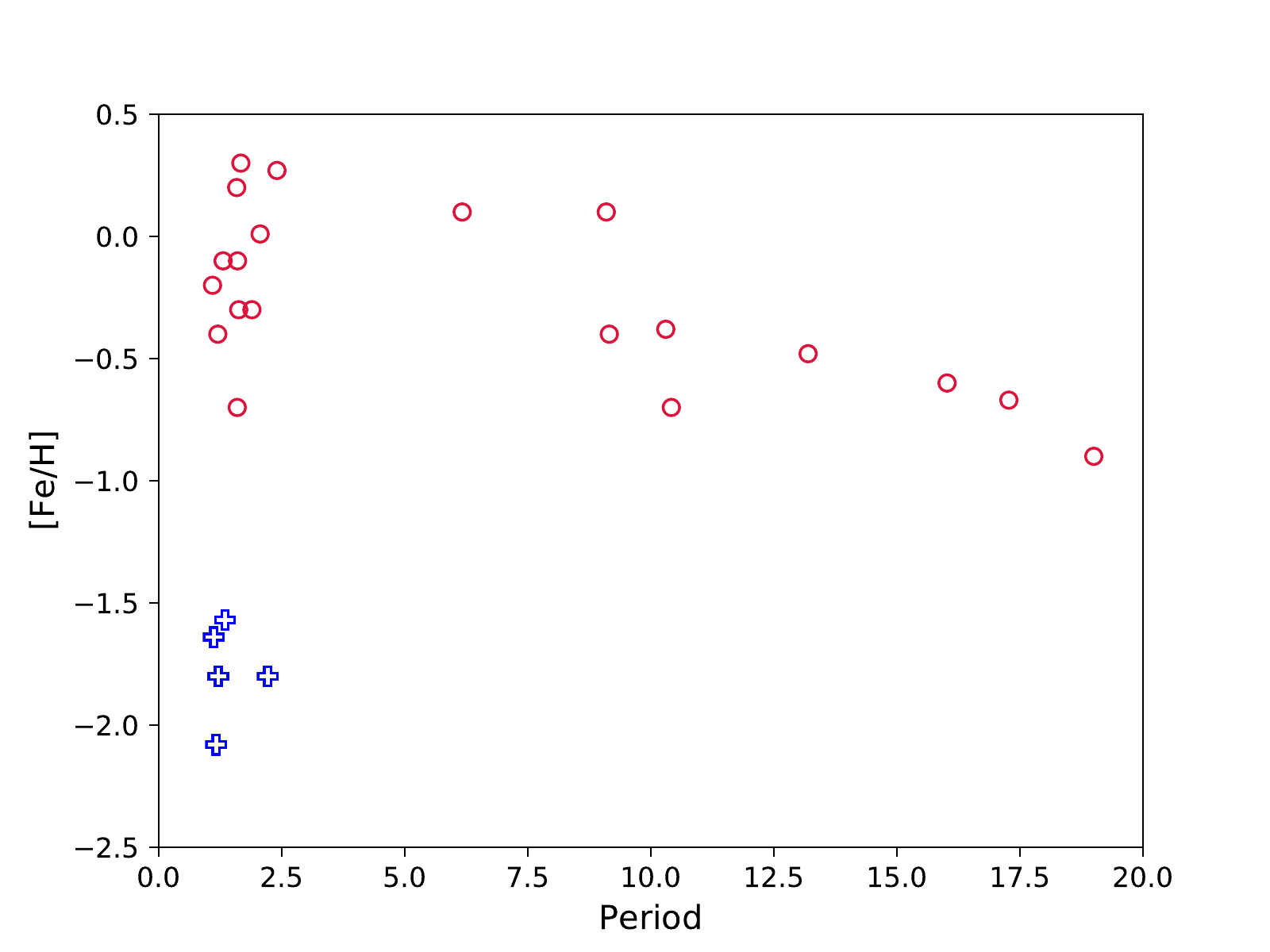}
\centering
\caption{Type II Cephieds in the Galactic Disk and Halo. \\ 
Disk Cephieds are represented by red markers, while Halo Cepheids are represented by blue markers. \\
They are seperated by [Fe/H] = --1.0.}
\label{fig:fieldceph}
\end{figure*}

The distances and proper motions of T2C stars had been largely unknown until {\it Gaia} DR2 (\citealt{Luri2018}). Hence, we have assembled a list of T2C field stars with their periods, distances, absolute magnitudes, and metallicities. Our assembled data for T2C stars in the galactic field are shown in Table~\ref{tab:atmparam}. We plotted [Fe/H] against period for each T2C star in the field in Figure~\ref{fig:fieldceph}, where we separate the halo and disk cepheids by [Fe/H] = --1.0.

For the globular clusters, data are to be found in the catalogue of  C.~Clement (updated January 2017), combined with the catalogues of H.~Sawyer Hogg (\citealt{Clement2001}), which includes the metallicities of each cluster with a T2C star. We have omitted the cluster Omega~Cen, and the 2 unusual globulars NGC~6388 and NGC~6441, as they are probably the remains of elliptical galaxies (\citealt{Corwin2006}).\footnote{Omega~Cen is a special environment for variable stars because it seems to contain stars with [Fe/H] from 0.0 to --2.5. According to \citet{Clement2001}, it has 9 stars classified as Cepheids, and one RV~Tau star. According to \citet{Gonzalez1994}, variables 1, 29, and 48 have [Fe/H] values of --1.77, --1.99, and --1.65 respectively. Two other stars that lie to the left of the red gaint branch, ROA 24 and 342 have [Fe/H] = --2.05 and --1.97. Thus, Omega~Cen, despite its very wide range in metallicity, shows only metal-poor variables, and similar stars that lie above, or to the left of, the red giant branch.} For the T2C stars in the Galactic Bulge, see the paper by \citet{Bhardwaj2017}. A few stars with periods over 20 days have been excluded because they are rare and may show RV~Tau behavior.\footnote{RV~Tau behavior is defined as alternating low and high minimum light. It is usally seen in T2C stars with periods greater than 20 days, however, the term has also been used for T2C stars with periods greater than 20 days, even though their minima have not shown alteration.}

\begin{figure*}
\includegraphics[width=0.75\textwidth]{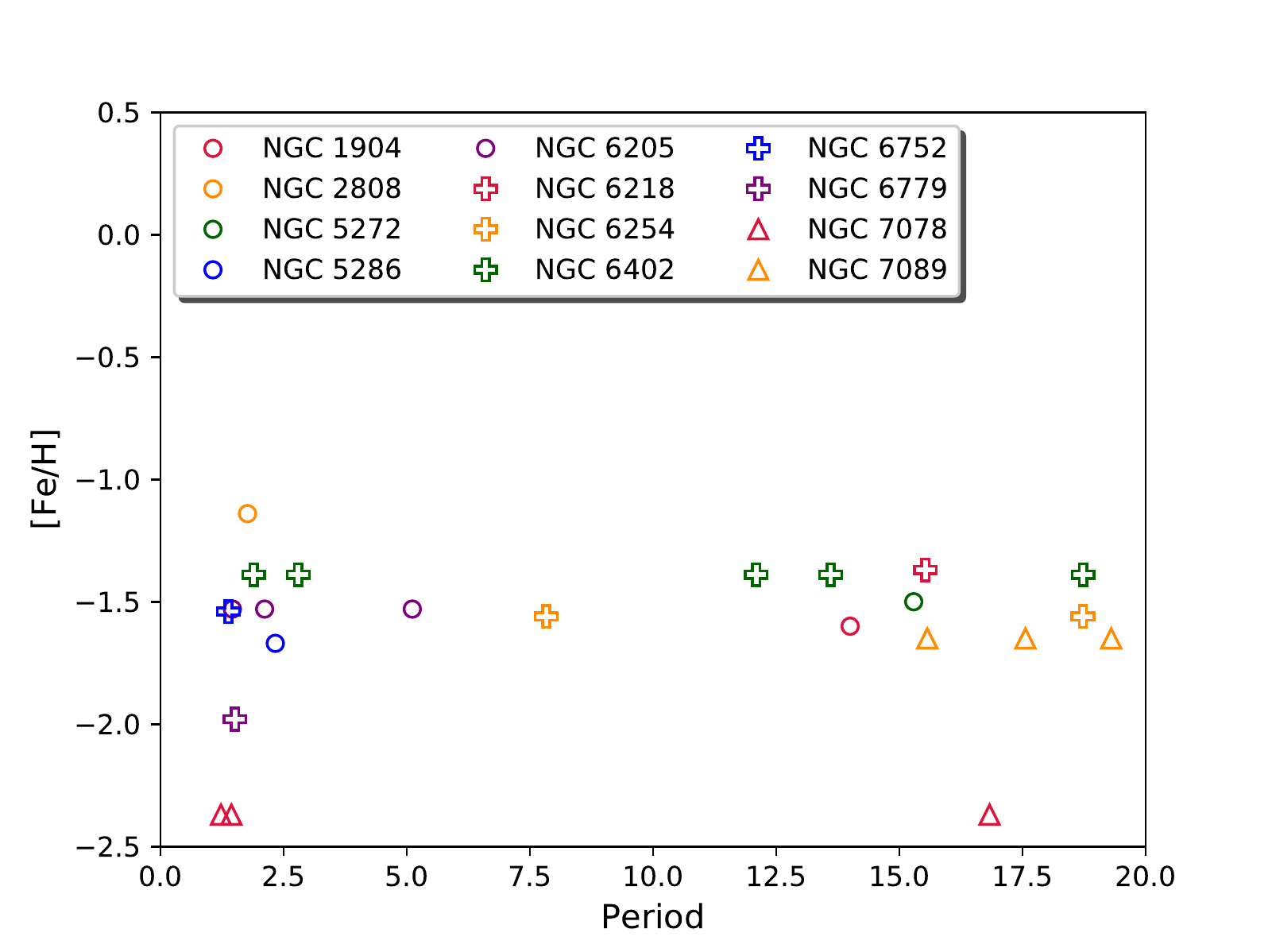}
\centering
\caption{Globular Cluster with Type II Cepheids}
\label{fig:glob}
\end{figure*}

To compare the field and globular cluster T2C stars, we plotted [Fe/H] against period for each star that is a cluster member in Figure~\ref{fig:glob}.

%table 1
\begin{table}
    \scriptsize
    \centering
    \caption{Field Type II Cepheids from \it Gaia}
    \label{tab:atmparam}
    {\begin{tabular}{lrrrrrr}
    \hline \hline Star & P(days) & $\pi$(mas) & M$_{G}$ & [Fe/H] \\ \hline
    
    BX~Del & 1.092 & 0.343 $\pm$ 0.039 & --0.132 & --0.20 \\
    V716~Oph & 1.116 & 0.359 $\pm$ 0.045 & --0.159 & --1.64 \\
    BF~Ser & 1.165 & 0.247 $\pm$ 0.036 & --0.925 & --2.08 \\
    VY~Pyx & 1.200 & 3.902 $\pm$ 0.053 & 0.036 & --0.40 \\
    CE~Her & 1.209 & 0.200 $\pm$ 0.036 & --0.973 & --1.80 \\
    BV~Cha & 1.238 & 0.365 $\pm$ 0.026 & --0.005 & -- \\
    BL~Her & 1.307 & 0.793 $\pm$ 0.031 & --0.312 & --0.10 \\
    XX~Vir & 1.348 & 0.312 $\pm$ 0.040 & --0.139 & --1.57 \\
    MQ~Aql & 1.481 & 0.095 $\pm$ 0.025 & --1.294 & -- \\
    KZ~Cen & 1.520 & 0.248 $\pm$ 0.036 & --0.607 & -- \\
    SW~Tau & 1.584 & 1.115 $\pm$ 0.051 & --0.194 & 0.20 \\
    V745~Oph & 1.595 & 0.190 $\pm$ 0.027 & --0.473 & --0.70 \\
    NW~Lyr & 1.601 & 0.319 $\pm$ 0.029 & --0.139 & --0.10 \\
    V971~Aql & 1.625 & 0.567 $\pm$ 0.053 & 0.579 & --0.30 \\
    DU~Ara & 1.641 & 0.418 $\pm$ 0.040 & 0.044 & -- \\
    VZ~Aql & 1.668 & 0.232 $\pm$ 0.027 & 0.170 & 0.30 \\
    V439~Oph & 1.893 & 0.488 $\pm$ 0.034 & 0.291 & --0.30 \\
    RT~TrA & 1.946 & 1.027 $\pm$ 0.091 & --0.322 & -- \\
    V553~Cen & 2.060 & 1.718 $\pm$ 0.076 & --0.551 & 0.01 \\
    UY~Eri\footnote{\scriptsize Despite the fact that UY~Eri has a probable error only 4.35 times its parallax, it has been included, as it is used to define a class of stars.} & 2.213 & 0.248 $\pm$ 0.057 & --1.771 & --1.80 \\
    AU~Peg & 2.402 & 1.674 $\pm$ 0.045 & 0.171 & 0.27 \\
    CN~CMa & 2.546 & 0.196 $\pm$ 0.025 & --0.295 & -- \\
    V351~Cep & 2.806 & 0.496 $\pm$ 0.039 & --2.383 & -- \\
    BE~Pup & 2.871 & 0.156 $\pm$ 0.024 & --0.764 & -- \\
    LN~Pav & 3.600 & 0.896 $\pm$ 0.022 & 3.309 & -- \\
    BD~Cas & 3.652 & 0.327 $\pm$ 0.028 & --1.960 & -- \\
    V383~Cyg & 4.612 & 0.475 $\pm$ 0.027 & --1.497 & -- \\
    V394~Cep & 5.689 & 0.105 $\pm$ 0.021 & --1.729 & -- \\
    TX~Del & 6.166 & 0.978 $\pm$ 0.040 & --0.988 & 0.10 \\    
    PZ~Aql & 8.756 & 0.525 $\pm$ 0.040 & --0.426 & -- \\
    V1043~Cyg & 8.847 & 0.214 $\pm$ 0.029 & --1.230 & -- \\    
    $\kappa$~Pav & 9.094 & 5.199 $\pm$ 0.309 & 0.336 & 0.10 \\
    IX~Cas & 9.155 & 0.222 $\pm$ 0.037 & --2.007 & --0.40 \\
    AL~Vir & 10.303 & 0.346 $\pm$ 0.047 & --2.918 & --0.38 \\
    AP~Her & 10.416 & 0.343 $\pm$ 0.038 & --1.709 & --0.70 \\
    CQ~Cha & 12.300 & 0.497 $\pm$ 0.018 & 1.468 & -- \\
    AL~Lyr & 12.992 & 0.228 $\pm$ 0.027 & --1.529 & -- \\
    DD~Vel & 13.195 & 0.454 $\pm$ 0.031 & 0.032 & --0.48 \\
    V2338~Oph & 13.700 & 0.260 $\pm$ 0.035 & --1.334 & -- \\
    FI~Sct & 14.862 & 0.168 $\pm$ 0.032 & --0.410 & -- \\    
    CO~Pup & 16.019 & 0.263 $\pm$ 0.027 & --2.123 & --0.60 \\    
    W~Vir & 17.274 & 0.399 $\pm$ 0.066 & --1.927 & --0.67 \\    
    ST~Pup & 19.000 & 0.371 $\pm$ 0.026 & --2.113 & --0.90 \\    
    CY~Vel & 19.529 & 0.174 $\pm$ 0.027 & --1.338 & -- \\
    RS~Pav & 19.954 & 0.353 $\pm$ 0.037 & --1.870 & -- \\ \hline
    
    \end{tabular}}
    
    %\begin{itemize}
    %\item[\it Remarks:] {\it }
    %\end{itemize}
        
\end{table}

Globular clusters show an absence of T2C  stars with periods between about 3 and 9 days, with the exception of variable~3 in M10, and variable~2 in M13, which fall in the middle of the gap for variables in globular clusters. The field stars show a deficiency of variables with [Fe/H] $\textless$ --1.0, with the exception of a small clump with periods between 1 and 3 days, and [Fe/H] between --1.5 and --2.0 recently analysed by \citet{Kovtyukh2018} and called UY~Eridani stars. Short period T2C stars with significantly larger metallicities are usually called BL~Herculis stars.

%figure 1

%figure 2

\section{Kinematics of T2C Stars in the Field}
\label{sec:kinfield}

Stellar populations are closely associated with the galactic orbits of the objects in question. Their stellar motions with respect to the local standard of rest (LSR) may best be parameterized by their U, V, W components of motion, where the component U is positive toward the Galactic Anticenter, V is positive in the direction of Galactic Rotation, and W is positive toward the North Galactic Pole. In Table~\ref{tab:cephparam}, we show the U, V, W, components for each T2C star with a parallax of at least 5 times its probable error in the {\it Gaia} catalogue. The U, V, W components of motion were calculated using the gal$\_$uvw script (\citealt{Koposov2016})\footnote{\scriptsize $\url{github.com/segasai/astrolibpy/blob/master/astrolib/gal\_uvw.py}$}  available on GitHub, which follows the outline of \citet{Johnson1987}, except that U is positive toward the Galactic Anticenter, rather than the Galactic Center, and the J2000 transformation matrix to galactic coordinates was taken from the Hipparcos and Tycho Catalogues (\citealt{ESA1997}). Radial velocities were taken from {\it Gaia} DR2, \citet{Beers2000}, \citet{Kordopatis2013}, \citet{Kharchenko2007}, and \citet{Barbier1994}. The radial velocities in {\it Gaia} DR2 are preliminary, because they were measured during each visit with the {\it Gaia} spectrograph, which covers a wavelength from 8500 -- 8900 \AA{} with a resolution of 11,000, just as the Radial Velocity Experiment (RAVE) did. In Figure~\ref{fig:vel}, we show the dependence of the U, V, and W components on the metallicity of each star. The distribution of the W components is the most important for distinguishing the population of halo stars. T2C stars with [Fe/H] from 0.0 to --1.0, show a distribution from --60 to 100 \kms, with single outliers near --100 and 180 \kms. For the metal-poor stars with [Fe/H] $\textless$ --1.5, the W components lie from --100 to 100. It is interesting to note the large negative V components for the metal-poor stars, indicating that their orbits are retrograde and the stars must have been captured. Despite the current century being called the ``era of big data'', there are so few stars in this category that little is proven from this sample.

%table 2
\begin{table*}[t]
    \scriptsize
    \centering
    \caption{Kinematics of Field Type II Cepheids}
    \label{tab:cephparam}
    {\begin{tabular}{lrrrrrrrr}
    \hline \hline Star & l(deg) & b(deg) & $\mu_{\alpha \ast}$(mas yr$^{-1}$) & $\mu_{\delta}$(mas yr$^{-1}$) & $v_{r}$(\kms) & U(\kms) & V(\kms) & W(\kms) \\ \hline
    
BX~Del & 59.939 & --10.281 & --1.142 $\pm$ 0.054 & --6.530 $\pm$ 0.054 & --6.964 $\pm$ 7.344 & --74.604$_{-09.952}^{+07.920}$ & --40.292$_{-06.672}^{+05.309}$ & --28.035$_{-04.351}^{+03.463}$ \\
V716~Oph & 9.857 & 27.795 & --25.789 $\pm$ 0.072 & --7.152 $\pm$ 0.039 & --230.000 $\pm$ 30.000 & 249.531$_{-08.319}^{+06.466}$ & --306.850$_{-42.032}^{+32.668}$ & 99.525$_{-29.199}^{+22.694}$ \\
BF~Ser & 22.865 & 54.822 & --15.362 $\pm$ 0.055 & --13.289 $\pm$ 0.069 & --146.084 $\pm$ 9.678 & 56.735$_{-02.471}^{+01.842}$ & --397.089$_{-66.379}^{+49.491}$ & --17.449$_{-16.575}^{+12.358}$ \\
VY~Pyx & 248.771 & 13.617 & 10.996 $\pm$ 0.067 & 29.424 $\pm$ 0.052 & 21.031$\pm$ 1.931 & 14.659$_{-00.226}^{+00.220}$ & 8.675$_{-00.175}^{+00.170}$ & 43.098$_{-00.334}^{+00.325}$ \\
CE~Her & 39.125 & 22.037 & --2.289 $\pm$ 0.055 & --3.359 $\pm$ 0.068 & --258.000 $\pm$ 30.000 & 122.473$_{-13.318}^{+09.255}$  & --215.201$_{-19.100}^{+13.273}$ & --73.779$_{-04.250}^{+02.953}$ \\
BV~Cha & 303.439 & --16.896 & --2.348 $\pm$ 0.053 & --1.322 $\pm$ 0.042 & --0.628 $\pm$ 4.089 & 20.419$_{-02.921}^{+02.532}$ & 0.507$_{-01.314}^{+01.139}$ & --8.448$_{-01.690}^{+01.465}$ \\
BL~Her & 45.161 & 19.463 & --4.140 $\pm$ 0.042 & --11.750 $\pm$ 0.060 & 19.031 $\pm$ 5.429 & --74.923$_{-02.294}^{+02.481}$ & --25.247$_{-02.243}^{+02.426}$ & 8.577$_{-00.077}^{+00.083}$ \\
XX~Vir & 337.777 & 50.717 & --11.460 $\pm$ 0.080 & --10.899 $\pm$ 0.070 & --55.000 $\pm$ 30.000 & 74.236$_{-06.086}^{+07.876}$ & --205.947$_{-27.799}^{+35.976}$ & --69.773$_{-03.993}^{+05.167}$ \\
MQ~Aql & 49.842 & --4.960 & --10.423 $\pm$ 0.038 & --7.598 $\pm$ 0.034 & -- & -- & -- & -- \\
KZ~Cen & 294.076 & 15.740 & 8.054 $\pm$ 0.049 & --3.898 $\pm$ 0.034 & 201.967 $\pm$ 3.588 & --243.761$_{-19.220}^{+25.747}$ & --107.048$_{-06.773}^{+09.073}$ & 20.362$_{-05.878}^{+07.875}$ \\
SW~Tau & 190.140 & --29.867 & 3.892 $\pm$ 0.109 & --8.803 $\pm$ 0.062 & 10.900 $\pm$ 0.300 & --10.031$_{-00.715}^{+00.784}$ & --27.180$_{-01.613}^{+01.767}$ & --5.588$_{-00.731}^{+00.801}$ \\
V745~Oph & 25.654 & 22.033 & --0.178 $\pm$ 0.046 & --10.354 $\pm$ 0.041 & -- & -- & -- & -- \\
NW~Lyr & 66.477 & 10.350 & 2.042 $\pm$ 0.044 & --3.372 $\pm$ 0.048 & --10.771 $\pm$ 7.991 & --36.973$_{-03.486}^{+04.184}$ & --0.933$_{-00.764}^{+00.917}$ & --43.831$_{-03.783}^{+04.539}$ \\
V971~Aql & 26.992 & --12.585 & --5.848 $\pm$ 0.091 & --0.047 $\pm$ 0.074 & --40.710 $\pm$ 10.377 & 8.615$_{-02.006}^{+02.419}$ & --19.481$_{-01.953}^{+02.355}$  & 58.212$_{-04.036}^{+04.869}$ \\
DU~Ara & 327.931 & --15.214 &  --1.923 $\pm$ 0.036 & --6.997 $\pm$ 0.043 & --21.591 $\pm$ 4.232 & 56.190$_{-04.393}^{+05.323}$ & --39.489$_{-06.082}^{+07.369}$ & --9.590$_{-01.815}^{+02.199}$ \\
VZ~Aql & 28.359 & --6.133 & --2.428 $\pm$ 0.049 & --4.457 $\pm$ 0.046 & 105.000 $\pm$ NaN & --149.926$_{-05.754}^{+07.270}$ & --28.068$_{-10.485}^{+13.247}$ & --1.308$_{-00.777}^{+00.982}$ \\
V439~Oph & 28.317 & 16.729 & --3.531 $\pm$ 0.053 & --11.534 $\pm$ 0.051 & --69.547 $\pm$ 8.315 & --9.844$_{-04.152}^{+04.774}$ & --116.874$_{-07.012}^{+08.062}$ & --33.495$_{-01.094}^{+01.258}$ \\
RT~TrA & 325.258 & --10.400 & --3.374 $\pm$ 0.098 & --11.447 $\pm$ 0.078 & --7.249 $\pm$ 6.765 & 29.248$_{-02.880}^{+03.440}$ & --21.122$_{-03.578}^{+04.274}$ & --15.227$_{-01.784}^{+02.131}$ \\
V553~Cen & 329.579 & 24.677 & 4.012 $\pm$ 0.164 & --1.689 $\pm$ 0.175 & --6.000 $\pm$ 2.700 & --11.771$_{-00.074}^{+00.081}$ & 20.231$_{-00.368}^{+00.402}$ & --4.075$_{-00.544}^{+00.594}$ \\
UY~Eri & 193.341 & --52.629 & 24.426 $\pm$ 0.076 & --6.348 $\pm$ 0.095 & 177.327 $\pm$ 8.515 & 274.845$_{-31.857}^{+50.871}$ & --410.894$_{-074.947}^{+119.681}$ & 68.693$_{-36.882}^{+58.895}$ \\
AU~Peg & 69.130 & --22.266 & --2.730 $\pm$ 0.070 & --12.887 $\pm$ 0.079 & 0.000 $\pm$ 4.400 & --36.584$_{-01.008}^{+01.063}$ & --4.636$_{-00.574}^{+00.606}$ & --10.191$_{-00.433}^{+00.457}$ \\
CN~CMa & 231.503 & --4.500 & --0.652 $\pm$ 0.036 & 1.708 $\pm$ 0.033 & -- & -- & -- & -- \\
V351~Cep & 105.151 & --0.719 & --3.352 $\pm$ 0.070 & --1.614 $\pm$ 0.063 & --21.214 $\pm$ 4.971 & --48.249$_{-03.283}^{+03.843}$ & 2.203$_{-00.889}^{+01.041}$ & 9.552$_{-00.034}^{+00.039}$ \\
BE~Pup & 240.555 & --2.997 & --1.252 $\pm$ 0.034 & 1.765 $\pm$ 0.040 & 116.229 $\pm$ 7.586 & 105.258$_{-07.099}^{+09.680}$ & --55.241$_{-04.145}^{+05.652}$ & --7.059$_{-02.287}^{+03.118}$ \\
LN~Pav & 325.613 & --28.320 & 2.246 $\pm$ 0.025 & 1.086 $\pm$ 0.034 & -- & -- & -- & -- \\
BD~Cas & 117.994 & --0.958 & --1.936 $\pm$ 0.044 & --0.923 $\pm$ 0.040 & --49.300 $\pm$ 0.300 & --58.064$_{-02.676}^{+03.177}$ & --16.263$_{-01.401}^{+01.663}$ & --1.383$_{-01.119}^{+01.328}$ \\
V383~Cyg & 73.927 & --2.765 & --0.620 $\pm$ 0.045 & --3.943 $\pm$ 0.042 & --24.400 $\pm$ 0.300 & --35.673$_{-02.374}^{+02.660}$ & --20.715$_{-00.727}^{+00.814}$ & --10.282$_{-00.852}^{+00.955}$ \\
V394~Cep & 102.939 & 3.299 & --2.348 $\pm$ 0.039 & --1.448 $\pm$ 0.037 & -- & -- & -- & -- \\
TX~Del & 50.958 & --24.263 & --2.989 $\pm$ 0.067 & --7.693 $\pm$ 0.046 & 13.594 $\pm$ 2.887 & --44.895$_{-01.432}^{+01.554}$ & --4.135$_{-01.237}^{+01.342}$ & --5.911$_{-00.130}^{+00.141}$ \\
PZ~Aql & 30.880 & --2.312 & 0.241 $\pm$ 0.104 & --7.587 $\pm$ 0.087 & --6.282 $\pm$ 5.744 & --32.770$_{-02.652}^{+03.090}$ & --42.037$_{-04.585}^{+05.341}$ & --26.331$_{-01.897}^{+02.210}$ \\
V1043~Cyg & 75.125 & 1.960 & --3.601 $\pm$ 0.045 & --6.295 $\pm$ 0.053 & --16.388 $\pm$ 3.563 & --159.334$_{-19.803}^{+26.011}$ & --43.266$_{-05.222}^{+06.859}$ & --4.316$_{-01.060}^{+01.392}$ \\
$\kappa$~Pav & 328.287 & --25.388 & --8.957 $\pm$ 0.325 & 15.091 $\pm$ 0.376 & 37.800 $\pm$ 0.800 & --47.734$_{-00.458}^{+00.516}$ & 3.401$_{-00.085}^{+00.096}$ & --0.294$_{-00.727}^{+00.819}$ \\
IX~Cas & 115.347 & --11.949 & --4.481 $\pm$ 0.045 & --1.980 $\pm$ 0.036 & --104.000 $\pm$ 2.900 & --146.204$_{-14.335}^{+20.068}$ & --39.470$_{-05.891}^{+08.248}$ & 4.555$_{-03.838}^{+05.373}$ \\
AL~Vir & 330.999 & 45.173 & --4.755 $\pm$ 0.084 & --0.072 $\pm$ 0.072 & 14.393 $\pm$ 4.327 & 28.179$_{-05.857}^{+07.698}$ & --34.011$_{-06.326}^{+08.314}$ & 35.833$_{-02.043}^{+02.685}$ \\
AP~Her & 47.090 & 7.404 & --2.325 $\pm$ 0.062 & --6.991 $\pm$ 0.069 & --31.865 $\pm$ 5.557 & --62.086$_{-08.276}^{+10.339}$ & --77.152$_{-07.507}^{+09.378}$ & --11.281$_{-01.052}^{+01.314}$ \\
CQ~Cha & 302.860 & --15.062 & --10.049 $\pm$ 0.031 & 4.580 $\pm$ 0.032 & -- & -- & -- & -- \\
AL~Lyr & 60.592 & 6.583 & --1.896 $\pm$ 0.036 & --1.267 $\pm$ 0.045 & --3.608 $\pm$ 4.952 & --41.563$_{-04.590}^{+05.823}$ & --12.398$_{-02.936}^{+03.724}$ & 28.925$_{-02.630}^{+03.337}$ \\
DD~Vel & 271.505 & --1.387 & --1.026 $\pm$ 0.054 & --1.857 $\pm$ 0.058 & 36.235 $\pm$ 3.346 & --16.196$_{-00.481}^{+00.552}$ & --22.144$_{-00.064}^{+00.073}$ & --15.477$_{-02.121}^{+02.432}$ \\
V2338~Oph & 34.847 & 13.786 & --3.866 $\pm$ 0.059 & --13.087 $\pm$ 0.059 & --166.727 $\pm$ 4.592 & --23.917$_{-18.235}^{+23.909}$ & --274.769$_{-24.310}^{+31.874}$ & --73.717$_{-04.380}^{+05.743}$ \\
FI~Sct & 26.502 & --1.858 & --1.225 $\pm$ 0.050 & --4.975 $\pm$ 0.048 & -- & -- & -- & -- \\
CO~Pup & 250.444 & 4.521 & --4.445 $\pm$ 0.037 & 5.822 $\pm$ 0.036 & 77.644 $\pm$ 5.122 & 141.862$_{-11.484}^{+14.112}$ & --15.775$_{-03.970}^{+04.878}$ & 7.559$_{-01.301}^{+01.599}$ \\
W~Vir & 319.566 & 58.371 & --3.944 $\pm$ 0.094 & 1.027 $\pm$ 0.102 & --59.017 $\pm$ 17.301 & 57.818$_{-06.349}^{+08.866}$ & 14.457$_{-04.024}^{+05.619}$ & --31.303$_{-01.369}^{+01.912}$ \\
ST~Pup & 246.852 & --16.492 & --0.660 $\pm$ 0.043 & 3.613 $\pm$ 0.047 & 45.000 $\pm$ 1.000 & 51.791$_{-02.458}^{+02.828}$ & --10.660$_{-01.073}^{+01.235}$ & 2.689$_{-00.070}^{+00.080}$ \\
CY~Vel & 273.397 & --3.9341 & --3.670 $\pm$ 0.047 & 6.256 $\pm$ 0.041 & 61.578 $\pm$ 5.557 & 180.574$_{-25.928}^{+35.453}$ & --62.229$_{-01.820}^{+02.489}$ & 43.540$_{-04.077}^{+05.575}$ \\
RS~Pav & 335.010 & --17.638 & --2.054 $\pm$ 0.058 & --0.761 $\pm$ 0.054 & 10.000 $\pm$ NaN & --14.346$_{-00.533}^{+00.657}$ & --11.842$_{-02.861}^{+03.531}$  & 23.664$_{-02.283}^{+02.817}$ \\ \hline
        
        \end{tabular}}

    \begin{itemize}
    \item[\it Remarks:] {\it Type II Cepheids are ordered by period, as they are in Table~\ref{tab:atmparam}.}
    \end{itemize}
    
\end{table*}

\section{Kinematics of T2C in Globular Clusters}
\label{sec:kingc}

\begin{figure*}
\includegraphics[width=1.0\textwidth]{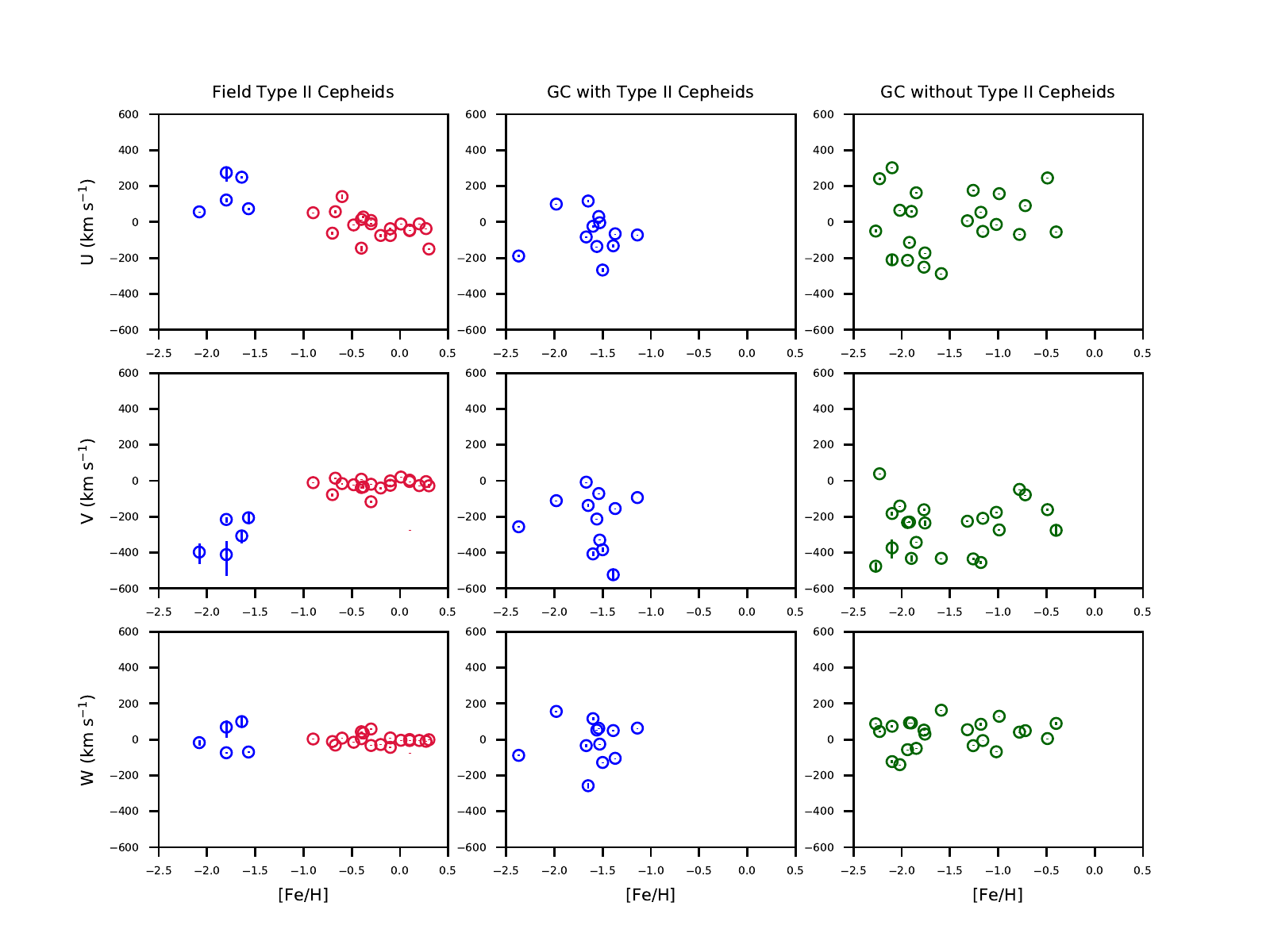}
\centering
\caption{U, V, W Components of T2C in the Galactic Field, Globular Clusters, and Globular Clusters without Cepheids}
\label{fig:vel}
\end{figure*}

We can still compare the small sample of T2C stars with globular clusters that contain Cepheids as shown on the right side of Figure~\ref{fig:vel}, where the data for their distances, proper motions, and radial velocities have been obtained from the \citet{Helmi2018} and \citet{Harris1996}.

In Table~\ref{tab:t2cglobparam}, we show the kinematics of globular clusters with at least 1 T2C star. The sample of T2C stars in globular clusters must be nearly complete due to the many investigations of variable stars in clusters. They are found in globulars with [Fe/H] $\leq$ --1.0.

There are other environments with T2C stars including the Large Magellanic Cloud (\citealt{Soszynski2008}), the Small Magellanic Cloud (\citealt{Soszynski2010}), the Fornax system (\citealt{Bersier2002}), and the Galactic Bulge (\citealt{Bhardwaj2017}), but none of them have been investigated for their metallicity.

In Table~\ref{tab:globparam}, we show the kinematics of globular clusters that lack T2C stars.

%figure 3

\section{Discussion}

An examination of Figure~\ref{fig:vel} permits the following relationships between T2C stars in the field and in globular clusters.

\small
\begin{itemize}
    \item[\it 1 --] {\it There are no globular clusters with [Fe/H] $\textgreater$ --1.0 and a T2C star (excluding
Omega~Cen, NGC~6388, and NGC~6441).}
    \item[\it 2 --] {\it The field T2C stars are rather well clumped in the U, V, and particularly W components.}
    \item[\it 3 --] {\it Globular clusters with T2C stars have widely spead U, V, and W velocities.}
    \item[\it 4 --] {\it Non-bulge globulars without Cepheids show a wide range in U and V components, and a significant range in W, which means a wide range in height above of below the Galactic Plane to which they penetrate. It was on this basis that the globulars have been used to define population II.}
    \item[\it 5 --] {\it As shown by their V components, the 5 halo T2C stars appear to be in retrograde orbits, indicating that they are captured.}
\end{itemize}

\section{Conclusions}
\label{sec:conc}

We confirm the suggestion by \citet{Woolley1966} that most of the T2C field stars of the Galactic Halo belong to a different population than the globular clusters of the Halo. This raises a number of questions as to the origin of the Halo.

The origin of the Galactic Halo has been an important aspect of research on galactic structure ever since the Vatican Conference (\citealt{OConnell1958}). Two different approaches have been suggested; the model of \citet{Eggen1962}, which suggests that the Halo consists of material left behind as the galaxy contracted to a central bulge with spiral arms, and that of \citet{Searle1978}, who proposed a model in which the Halo consists of small systems that have been captured over time by the galaxy's gravitational field. The latter is supported by the discovery of fragments
such as those described by \citet{Willman2010} and many others over the past decade (\citealt{McConnachie2012}), most recently, \citet{Myeong2018}.

Of the various stellar types found in the Halo, the T2C stars have the highest visual luminosity, and are easy to recognize by their periods from 1 to 20 days. However, it is surprising that the 2 elliptical companions to M31, NGC~147 and NGC~185, do not have any T2C stars at all \citep{Monelli2017} despite the presence of many RR Lyrae stars in both galaxies. In addition, a search for variables in 6 small companion galaxies of M31 revealed 870 RR~Lyrae stars, and 15 annomalous Cepheids, but not a single recognized T2C star \citep{Martinez2017}. A search for T2C stars in the giant ellipticals in nearby galaxy clusters would require photometry down to V=28, and probably a telecope of 30-m aperture. In the meantime, we await the discovery of additional T2C stars in our galaxy with {\it Gaia}, Pan-STARRS, and the Large Synoptic Survey Telescope (LSST).

\
 
This work has made use of data from the European Space Agency (ESA) mission {\it Gaia} (\url{https://www.cosmos.esa.int/gaia}), processed by the {\it Gaia} Data Processing and Analysis Consortium (DPAC,
\url{https://www.cosmos.esa.int/web/gaia/dpac/consortium}). Funding for the DPAC has been provided by national institutions, in particular the institutions participating in the {\it Gaia} Multilateral Agreement.

\

This research has been supported by the Kennilworth Fund of the New York Community Trust. George Wallerstein thanks Halton C. Arp for introducing him to the T2C stars in 1954. We also greatly appreciate the comments by Julie Lutz, Charli Sakari, Andrea Kunder, Dana Casetti-Dinescu, and Donald Serna-Grey on a preliminary draft of this manuscript.

\clearpage

%table 3
\begin{sidewaystable}
    \scriptsize
    \centering
    \caption{Kinematics of Globular Clusters with Type II Cepheids}
    \label{tab:t2cglobparam}
    {\begin{tabular}{lrrrrrrrrrrr}
    \hline \hline Cluster & [Fe/H] & l(deg) & b(deg) & $\pi$(mas) & $\mu_{\alpha \ast}$(mas yr$^{-1}$) & $\mu_{\delta}$(mas yr$^{-1}$) & $v_{r}$(\kms) & U(\kms) & V(\kms) & W(\kms) \\ \hline
    
    NGC~1904 & --1.60 & 227.229 & --29.352 & 0.036 $\pm$ 0.002 & 2.470 $\pm$ 0.005 & --1.560 $\pm$ 0.005 & 206.430 $\pm$ 0.870 & --23.722$_{-06.754}^{+07.420}$ & --407.100$_{-12.959}^{+14.236}$ & 115.874$_{-08.761}^{+09.625}$ \\
    NGC~2808 & --1.14 & 282.193 & --11.253 & 0.056 $\pm$ 0.001 & 1.003 $\pm$ 0.003 & 0.279 $\pm$ 0.003 & 104.610 $\pm$ 1.260 & --71.967$_{-00.446}^{+00.456}$ & --93.611$_{-00.005}^{+00.005}$ & 63.373$_{-00.448}^{+00.458}$ \\
    NGC~5272 & --1.50 & 42.217 & 78.707 & 0.027 $\pm$ 0.001 & --0.113 $\pm$ 0.003 & --2.627 $\pm$ 0.002 & --146.480 $\pm$ 0.660 & --267.127$_{-10.033}^{+10.820}$ & --383.906$_{-14.348}^{+15.473}$ & --127.833$_{-00.441}^{+00.476}$ \\
    NGC~5286 & --1.67 & 311.614 & 10.568 & 0.017 $\pm$ 0.003 & 0.184 $\pm$ 0.008 & --0.148 $\pm$ 0.007 & 56.800 $\pm$ 1.660 & --83.016$_{-03.367}^{+04.544}$ & --7.833$_{-01.041}^{+01.406}$ & --34.063$_{-07.811}^{+10.542}$ \\
    NGC~6205 & --1.53 & 59.009 & 40.912 & 0.080 $\pm$ 0.001 & --3.176 $\pm$ 0.003 & --2.588 $\pm$ 0.003 & --245.620 $\pm$ 0.940 & --3.778$_{-00.905}^{+00.921}$ & --330.195$_{-01.775}^{+01.806}$ & --25.896$_{-01.217}^{+01.239}$ \\
    NGC~6218 & --1.37 & 15.715 & 26.313 & 0.156 $\pm$ 0.001 & --0.158 $\pm$ 0.004 & --6.768 $\pm$ 0.003 & --41.000 $\pm$ 0.510 & --65.411$_{-00.772}^{+00.785}$ & --154.608$_{-01.438}^{+01.462}$ & --104.793$_{-00.715}^{+00.727}$ \\
    NGC~6254 & --1.56 & 15.137 & 23.076 & 0.151 $\pm$ 0.001 & --4.703 $\pm$ 0.004 & --6.529 $\pm$ 0.003 & 76.760 $\pm$ 0.590 & --135.877$_{-00.557}^{+00.567}$ & --213.050$_{-02.385}^{+02.429}$ & 52.518$_{-00.200}^{+00.204}$ \\
    NGC~6402 & --1.39 & 21.324 & 14.804 & 0.054 $\pm$ 0.003 & --3.615 $\pm$ 0.007 & --5.036 $\pm$ 0.006 & --66.100 $\pm$ 1.800 & --132.245$_{-08.988}^{+09.941}$ & --523.037$_{-25.294}^{+27.978}$ & 49.607$_{-03.120}^{+03.451}$ \\
    NGC~6752 & --1.54 & 336.493 & --25.628 & 0.231 $\pm$ 0.001 & --3.191 $\pm$ 0.002 & --4.035 $\pm$ 0.002 & --26.120 $\pm$ 0.510 & 30.680$_{-00.091}^{+00.091}$ & --71.282$_{-00.494}^{+00.498}$ & 62.372$_{-00.237}^{+00.240}$ \\
    NGC~6779 & --1.98 & 62.659 & 8.336 & 0.070 $\pm$ 0.002 & --2.009 $\pm$ 0.005 & 1.65 $\pm$ 0.006 & --136.670 $\pm$ 1.000 & 99.811$_{-00.547}^{+00.571}$ & --110.823$_{-00.324}^{+00.338}$ & 156.236$_{-03.682}^{+03.842}$ \\
    NGC~7078 & --2.37 & 65.013 & --27.313 & 0.057 $\pm$ 0.001 & --0.624 $\pm$ 0.004 & --3.796 $\pm$ 0.004 & --105.580 $\pm$ 1.450 & --189.127$_{-05.724}^{+06.013}$ & --256.021$_{-04.605}^{+04.838}$ & --88.500$_{-03.401}^{+03.573}$ \\
    NGC~7089 & --1.65 & 53.371 & --35.770 & 0.059 $\pm$ 0.004 & 3.491 $\pm$ 0.008 & --2.150 $\pm$ 0.007 & --4.790 $\pm$ 0.270 & 116.961$_{-06.201}^{+06.981}$ & --137.525$_{-08.633}^{+09.720}$ & --257.455$_{-14.760}^{+16.618}$ \\ \hline
        
		\end{tabular}}
        
    %\begin{itemize}
    %\item[\it Remarks:] {\it }
    %\end{itemize}
   
\end{sidewaystable}
   
%table 4
\begin{sidewaystable}
    \scriptsize
    \centering
    \caption{Kinematics of Globular Clusters without Type II Cepheids}
    \label{tab:globparam}
    {\begin{tabular}{lrrrrrrrrrr}
    \hline \hline Cluster & [Fe/H] & l(deg) & b(deg) & $\pi$(mas) & $\mu_{\alpha \ast}$(mas yr$^{-1}$) & $\mu_{\delta}$(mas yr$^{-1}$) & $v_{r}$(\kms) & U(\kms) & V(\kms) & W(\kms) \\ \hline
    
		NGC~0104 & --0.72 & 305.895 & --72.082 & 0.196 $\pm$ 0.000 & 5.248 $\pm$	 0.002 & --2.519 $\pm$ 0.002 & --18.950 $\pm$ 0.420 & 91.394$_{-00.050}^{+00.050}$ & --78.301$_{-00.109}^{+00.109}$ & 49.102$_{-00.059}^{+00.059}$ \\
		NGC~0288 & --1.32 & 151.285 & --89.380 & 0.140 $\pm$ 0.002 & 4.239 $\pm$ 0.004 & --5.647 $\pm$ 0.003 & --49.060 $\pm$ 0.320 & 6.920$_{-00.092}^{+00.089}$ & --225.250$_{-03.637}^{+03.529}$ & 54.469$_{-00.018}^{+00.017}$ \\
		NGC~0362 & --1.26 & 301.533 & --46.247 & 0.079 $\pm$ 0.001 & 6.695 $\pm$ 0.005 & --2.518 $\pm$ 0.003 & 226.930 $\pm$ 0.770 & 176.241$_{-03.809}^{+03.695}$ & --434.817$_{-04.828}^{+04.683}$ & --34.436$_{-02.033}^{+01.972}$ \\
		NGC~1851 & --1.18 & 244.513 & --35.036 & 0.030 $\pm$ 0.001 & 2.131 $\pm$ 0.004 & --0.622 $\pm$ 0.004 & 323.360 $\pm$ 1.040 & 54.251$_{-02.648}^{+02.460}$ & --454.826$_{-08.669}^{+08.052}$ & 84.532$_{-09.536}^{+08.857}$ \\
		NGC~2298 & --1.92 & 245.629 & --16.006 & 0.079 $\pm$ 0.002 & 3.276 $\pm$ 0.006 & --2.191 $\pm$ 0.006 & 147.410 $\pm$ 1.400 & --113.330$_{-04.263}^{+04.063}$ & --229.676$_{-02.770}^{+02.640}$ & 92.866$_{-02.662}^{+02.537}$\\
		NGC~3201 & --1.59 & 277.229 & 8.640 & 0.172 $\pm$ 0.001 & 8.334 $\pm$ 0.002 & --1.990 $\pm$ 0.002 & 494.620 $\pm$ 0.370 & --287.201$_{-00.743}^{+00.738}$ & --431.642$_{-00.126}^{+00.125}$ & 162.828$_{-00.209}^{+00.208}$ \\
		NGC~4590 & --2.23 & 299.626 & 36.051 & 0.066 $\pm$ 0.003 & --2.764 $\pm$ 0.005 & 1.792 $\pm$ 0.004 & --95.200 $\pm$ 0.400 & 241.037$_{-08.491}^{+07.875}$ & 38.792$_{-01.957}^{+01.815}$  & 44.535$_{-03.430}^{+03.181}$  \\
		NGC~4833 & --1.85 & 303.604 & --8.015 & 0.116 $\pm$ 0.001 & --8.315 $\pm$ 0.004 & --0.937 $\pm$ 0.003 & 207.860 $\pm$ 0.570 & 162.823$_{-02.609}^{+02.565}$ & --343.040$_{-01.675}^{+01.647}$ & --49.533$_{-00.348}^{+00.342}$ \\
		NGC~5024 & --2.10 & 332.963 & 79.764 & 0.014 $\pm$ 0.002 & --0.147 $\pm$ 0.005 & --1.351 $\pm$ 0.003 & --64.330 $\pm$ NaN & --210.222$_{-29.808}^{+23.143}$ & --373.543$_{-58.437}^{+45.370}$ & --123.105$_{-09.594}^{+07.449}$ \\
		NGC~5897 & --1.90 & 342.946 & 30.294 & 0.068 $\pm$ 0.003 & --5.411 $\pm$ 0.005 & --3.460 $\pm$ 0.005 & 99.920 $\pm$ 1.310 & 59.398$_{-06.115}^{+05.665}$ & --432.159$_{-17.191}^{+15.925}$ & 92.010$_{-01.377}^{+01.276}$ \\
		NGC~5927 & --0.49 & 326.604 & 4.860 & 0.100 $\pm$ 0.002 & --5.047 $\pm$ 0.006 & --3.233 $\pm$ 0.006 & --115.700 $\pm$ 3.100 & 245.275$_{-03.604}^{+03.456}$ & --160.847$_{-05.449}^{+05.224}$ & 5.050$_{-00.124}^{+00.119}$ \\
		NGC~6121 & --1.16 & 350.973 & 15.972 & 0.500 $\pm$ 0.001 & --12.496 $\pm$ 0.003 & --18.979 $\pm$ 0.003 & 71.400 $\pm$ 0.300 & --51.995$_{-00.042}^{+00.042}$ & --208.979$_{-00.338}^{+00.337}$ & --6.028$_{-00.041}^{+00.041}$ \\
		NGC~6144 & --1.76 & 351.929 & 15.701 & 0.067 $\pm$ 0.004 & --1.765 $\pm$ 0.009 & --2.637 $\pm$ 0.006 & 195.850 $\pm$ 0.900 & --172.066$_{-01.626}^{+01.442}$ & --235.134$_{-14.909}^{+13.225}$ & 29.886$_{-01.727}^{+01.532}$ \\
		NGC~6171 & --1.02 & 3.373 & 23.011 & 0.148 $\pm$ 0.003 & --1.936 $\pm$ 0.006 & --5.949 $\pm$ 0.005 & --35.010 $\pm$ 0.890 & --13.350$_{-00.649}^{+00.627}$ & --175.721$_{-03.600}^{+03.475}$ & --68.295$_{-01.027}^{+00.992}$  \\
		NGC~6287 & --2.10 & 0.132 & 11.023 & 0.107 $\pm$ 0.005 & --4.887 $\pm$ 0.012 & --1.921 $\pm$ 0.008 & --292.450 $\pm$ 0.810 & 302.088$_{-01.164}^{+01.063}$ & --182.244$_{-09.950}^{+09.082}$ & 73.588$_{-06.090}^{+05.558}$ \\
		NGC~6333 & --1.77 & 5.544 & 10.705 & 0.093 $\pm$ 0.003 & --2.203 $\pm$ 0.008 & --3.208 $\pm$ 0.006 & 229.100 $\pm$ 7.000 & --251.065$_{-00.565}^{+00.533}$ & --161.491$_{-06.315}^{+5.960}$  & 52.160$_{-00.255}^{+00.241}$  \\
		NGC~6356 & --0.40 & 6.724 & 10.220 & 0.079 $\pm$ 0.007 & --3.768 $\pm$ 0.010 & --3.375 $\pm$ 0.006 & 27.000 $\pm$ 4.300 & --55.161$_{-01.873}^{+01.584}$ & --275.756$_{-27.293}^{+23.089}$ & 89.281$_{-07.398}^{+06.258}$ \\
		NGC~6362 & --0.99 & 325.555 & --17.570 & 0.097 $\pm$ 0.001 & --5.501 $\pm$ 0.003 & --4.742 $\pm$ 0.003 & --13.000 $\pm$ 0.600 & 157.839$_{-01.887}^{+01.845}$ & --273.684$_{-03.535}^{+03.456}$ & 129.152$_{-01.400}^{+01.368}$ \\
		NGC~6397 & --2.02 & 338.165 & --11.960 & 0.378 $\pm$ 0.001 & 3.291 $\pm$ 0.003 & --17.591 $\pm$ 0.003 & 19.180 $\pm$ 0.460 & 65.571$_{-00.183}^{+00.183}$ & --140.694$_{-00.314}^{+00.313}$ & --140.090$_{-00.252}^{+00.251}$ \\
		NGC~6809 & --1.94 & 8.793 & --23.272 & 0.171 $\pm$ 0.001 & --3.402 $\pm$ 0.003 & --9.264 $\pm$ 0.003 & 176.460 $\pm$ 0.570 & --213.306$_{-00.327}^{+00.323}$ & --232.168$_{-01.851}^{+01.827}$ & --56.858$_{-00.094}^{+00.093}$ \\
		NGC~6838 & --0.78 & 56.746 & --4.564 & 0.225 $\pm$ 0.001 & --3.384 $\pm$ 0.003 & --2.653 $\pm$ 0.003 & --21.010 $\pm$ 0.530 & --69.004$_{-00.389}^{+00.385}$ & --48.222$_{-00.239}^{+00.237}$ & 40.826$_{-00.165}^{+00.163}$ \\
		NGC~7099 & --2.27 & 27.179 & --46.836 & 0.075 $\pm$ 0.004 & --0.702 $\pm$ 0.006 & --7.222 $\pm$ 0.006 & --186.480 $\pm$ 0.930 & --50.296$_{-09.213}^{+08.275}$ & --475.539$_{-24.714}^{+22.199}$ & 87.567$_{-02.901}^{+02.606}$ \\ \hline
       
		\end{tabular}}
        
    %\begin{itemize}
    %\item[\it Remarks:] {\it }
    %\end{itemize}
    
\end{sidewaystable}

\clearpage

\end{document}